\documentclass[twocolumn,showpacs,prl]{revtex4-1}

\usepackage{amsmath}
\usepackage{amssymb}
\usepackage{graphicx}
\usepackage{dcolumn}
\usepackage{bm}
\newcommand{\sgn}{\mathop{\mathrm{sgn}}}
\newcommand{\ef}{\mathop{\varepsilon_{F}}}
\newcommand{\e}{\mathop{\varepsilon}}

\begin{document}

\draft

\title {Electron-phonon bound state in graphene}%: fine structure induced by a weak electron-phonon coupling}

\author{S. M. Badalyan}
\email{Samvel.Badalyan@ua.ac.be}
\affiliation{Department of Physics, University of Antwerp, Groenenborgerlaan 171, B-2020 Antwerpen, Belgium}
\author{F. M. Peeters}
\affiliation{Department of Physics, University of Antwerp, Groenenborgerlaan 171, B-2020 Antwerpen, Belgium}

\begin{abstract}
The fine structure of the Dirac energy spectrum in graphene induced by electron-optical phonon coupling is investigated in the portion of the spectrum near the phonon emission threshold. The derived new dispersion equation in the immediate neighborhood below the phonon threshold corresponds to an electron-phonon bound state. We find that the singular vertex corrections beyond perturbation theory increase strongly the electron-phonon binding energy. The predicted enhancement of the effective electron-phonon coupling can be measured using angle-resolved spectroscopy.
\end{abstract}

%\pacs{72.25.Dc, 72.10.-d, 73.63.Hs, 73.21.Fg}
\maketitle

%\paragraph{Introduction}
Discovery of graphene \cite{graphene,geimnovo} with its unique conical gapless bandstructure provides a new rich area for investigations of many-body physics of chiral massless fermions \cite{kotov}. Active theoretical efforts are directed to the study of the interaction effects of Dirac carriers with elementary excitations of charge density waves \cite{hwang1,gangad} and lattice vibrations \cite{piscanec,tse1,gonzalez}, which result in the velocity renormalization of the bare Dirac spectrum \cite{park,calandra,tse2,aleiner,benfatto} and in the formation of new quasiparticles such as plasmarons \cite{polini,plasmaron}, polarons \cite{stauber,bruder}, and plasmon-phonon complexes \cite{jablan}. The spectral and damping properties of these quasiparticles have been studied by means of such powerful experimental tools as Raman \cite{pinczuk,Pisana2007} and angle-resolved photoemission \cite{bostwick1} spectroscopies. These high precision measurements indicate significant modifications of the peculiar graphene bandstructure, induced by electron-electron and electron-phonon interactions.

The aim of the present paper is to investigate how the Dirac spectrum in graphene is modified near the threshold of optical phonon emission. Previous studies of this problem in bulk semiconductors \cite{levinsonrashba} have shown that even for weak electron-phonon coupling, $\alpha$, the branches of the true spectrum can be classified into three main groups, according to the effective average number of phonon states, $N=1-Z$, bound to the electron and forming complex quasiparticles. Here $Z=(1-\partial \Sigma(\e,p)/\partial\e)^{-1}|_{\e=\e(p)}$ with $\Sigma(\e,p)$ the electron mass operator and $\e(p)$ the energy of the electron-phonon complex with momentum $p$. The first group includes states with $N\sim\alpha^{2}$. The spectrum of these branches differs from the bare one by a simple renormalization of the energy of the order of $\alpha^{2}$. The second and third groups are the hybrid and bound states of electrons and phonons. The universal threshold nonanalyticities \cite{landau} are responsible for the existence of these electron-phonon complexes, respectively, with $N\sim 1/2$ and $N\sim 1$. Note that for the formation of the hybrid states, a resonance situation in the bare spectrum is required. 

The character of the threshold singularities depends on the competition between the kinetic energy of the electrons and phonons and their interaction energy, therefore, is largely determined by the dimensionality of the system. In graphene because of its two-dimensional nature the size-extend of electrons and phonons is strongly reduced and one can expect that the threshold nonanalyticities will result in strong modifications of the bare Dirac spectrum. Such a strong enhancement of the electron-phonon effective coupling and the binding energies of complex quasiparticles has been previously found in two dimensional semiconductor structures \cite{smb1988,fp}.
 
\begin{figure}[t]
\includegraphics[height=1.25cm]{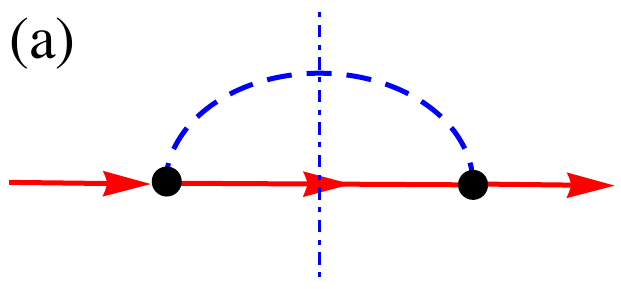}\hspace{.5cm}\includegraphics[height=1.25cm]{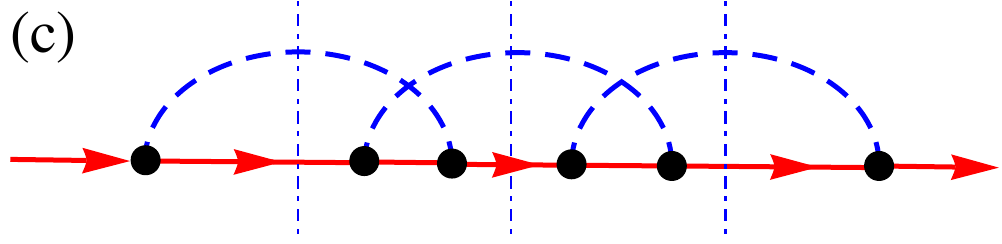}
\includegraphics[width=3cm]{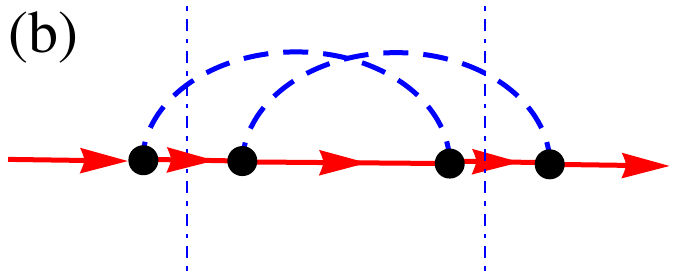}\hspace{.5cm}\includegraphics[height=1.25cm]{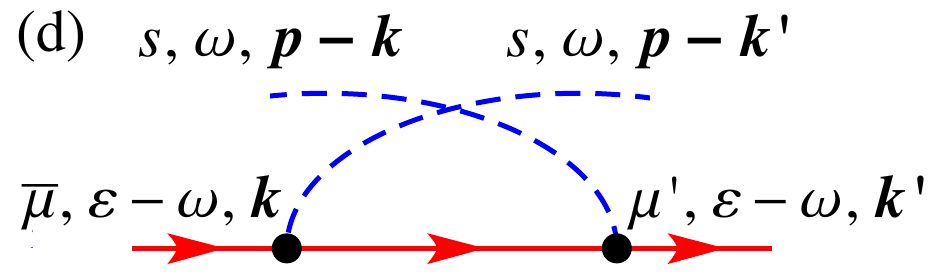}
\caption{The simplest diagrams for the electron self energy: (a) the lowest order mass operator $\Sigma_{0}(\e)$, (b) and (c) are vertex corrections to $\Sigma_{0}(\e)$ in next two orders. %The thin solid and dashed lines correspond to the bare electron and phonon Green functions, the dots to the bare electron-phonon vertices. 
The vertical dash-dotted lines show the dangerous intersections along one phonon and one electron lines corresponding to the threshold singularities of the mass operator. (d) The simplest diagram for the electron-phonon four vertex part from Fig.~\ref{fg2}.  }
\label{fg1}
\end{figure}

\begin{figure}[t]
\includegraphics[width=5.5cm]{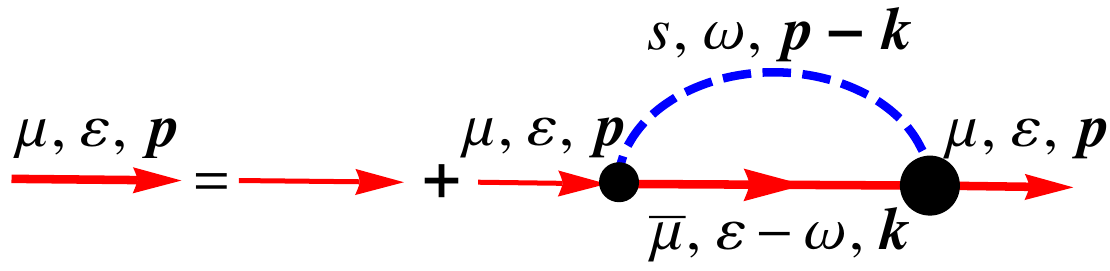}
\includegraphics[width=8.5cm]{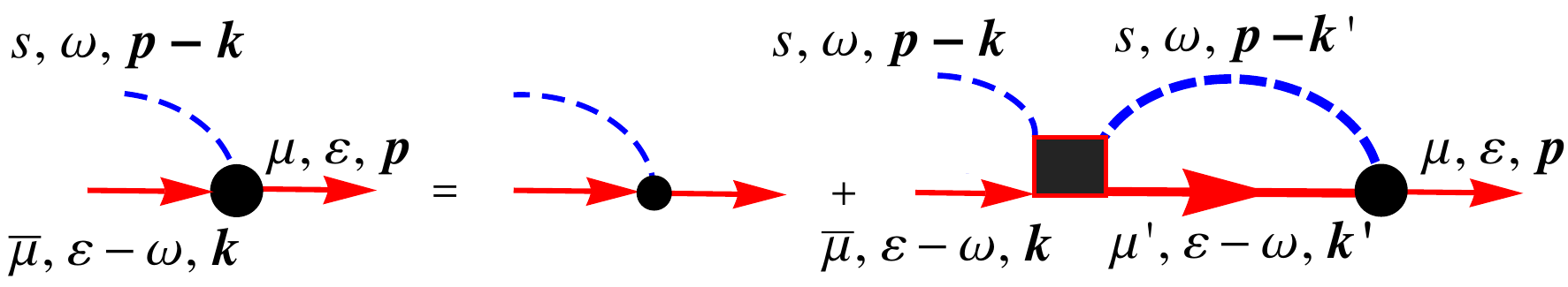}
\caption{(top) Dyson equation for the exact electron Green function $G_{\mu}(\e)$ (thick solid lines). (bottom) The ladder type equation for the exact electron-phonon vertex part $\Gamma_{\bar{\mu}\mu}(\e)$ (large bold dots). 
%The the thick and thin solid lines correspond to the exact and noninteracting electron Green functions, the large and small bold dots to the exact and bare electron-phonon vertices, and 
The solid square represents the irreducible four vertex part ${\bf \square}(\e)$ with two electron and two phonon external lines. The simplest diagram corresponding to ${\bf \square}(\e)$ is drawn in Fig.~\ref{fg1}(d).}
\label{fg2}
\end{figure}

Recent perturbative calculations \cite{park,calandra,tse2,stauber} of the electron mass operator, $\Sigma(\e)$, in graphene showed that in $n$-doped samples the real part of the lowest order contribution, $\Sigma_{0}(\e)$, diverges logarithmically at energies near the optical phonon emission threshold, $\e_{c}=\pm\omega_{0}$, while the imaginary part of $\Sigma_{0}(\e)$, related to the single-particle density of states, makes a discontinuous jump at the same threshold points. The $+$ ($-$) sign refers to the phonon emission process by Fermi electrons (holes), $\omega_{0}=196$ meV is the longitudinal optical phonon energy. We will use $\hbar=1$ units.
%As far as for weak coupling, perturbation theory is valid when the lowest order term is finite, 
Due to this non-analyticity the effective electron-phonon coupling becomes large and a perturbative calculation of $\Sigma_{0}(\e)$ in the neighborhood of the phonon emission threshold will not be a good approximation for $\Sigma(\e)$. In terms of the diagrammatic expansion, the diagrams with dangerous intersections along one phonon and one electron lines are responsible for the threshold singularities \cite{pit1959}. The simplest diagrams for $\Sigma(\e)$ with such dangerous intersections are shown in  Figs.~\ref{fg1}(a-c). As seen, the number of dangerous intersections increases with the order of the diagram and perturbation theory fails to converge when $\e  \rightarrow \e_{c}$. In order to find the true spectrum of the system in the neighborhood of $\e_{c}$, it is required to sum an infinite number of divergent diagrams with such dangerous intersections. For this we exploit an approach beyond perturbation theory \cite{levinsonrashba}, which leads to coupled integral equations for the exact electron Green function and for the exact electron-phonon vertex part, drawn in  Fig.~\ref{fg2}. By solving these equations we find a  new dispersion equation for the complex quasiparticle energy that in the immediate neighborhood below the threshold describes an electron-phonon bound state in graphene with new analytical dependencies on the electron-phonon coupling. The true spectrum does not asymptotically tend to the phonon energy but always remains below it at a small but finite distance. At finite wave vectors we find that the {\it singular vertex corrections} to $\Sigma_{0}(\e)$ increase strongly the electron-phonon binding energy in comparison with that calculated within the perturbative approach.

We search for new branches of electron-phonon complex states from the poles of the exact single-electron Green function, $G_{\mu}(\e)$, in the total energy parameter $\e$. Usually in doped graphene samples the Fermi energy, $\e_{F}$, is much larger than the lattice temperature $T$ and even at room temperatures most of the electrons are below $\e_{F}$. In the chiral basis the zero temperature Green function of noninteracting electrons
\begin{eqnarray}
G_{0\mu}(\e,{\bf k})=\frac{1}{\e+\e_{F}-\e_{\mu{\bf k}}+i 0 \cdot \sgn{\e}}
\label{egf}
\end{eqnarray}
corresponds to the thin solid lines in Figs.~\ref{fg1} and \ref{fg2}. The electron bare energy in the vicinity of the Dirac points in graphene has a linear dispersion, $\e_{\mu{\bf k}}=\mu v_{F} |\vec{k}|$, where $\vec{k}$ and $v_{F}$ are the momentum and the Fermi velocity of the massless fermions, described by the Hamiltonian $H_{0}=-v_{F} \vec{\sigma} \cdot \vec{k}$. The Pauli matrices $\vec{ \sigma}$ act in the pseudospace of graphene sublattices and $\mu=\pm1$ labels the electron chirality.
At low $T$ absorption of phonons by electrons is negligible and we replace all exact phonon Green functions (dashed lines) by free phonon propagators, $D_{s}(\omega,{\bf q})=(\omega-\omega_{s {\bf q}}+i 0)^{-1}-(\omega+\omega_{s {\bf q}}-i 0)^{-1}$,
%\begin{eqnarray}D_{s}(\omega,{\bf q})=\frac{1}{\omega-\omega_{s {\bf q}}+i 0}-\frac{1}{\omega+\omega_{s {\bf q}}-i 0}\end{eqnarray}
where $\omega_{s {\bf q}}$ and ${\bf q}$ are the energy and momentum of the $s$  phonon mode in graphene. 

The small dots in Figs.~\ref{fg1} and \ref{fg2} correspond to the bare electron-phonon vertex functions
\begin{eqnarray}
\gamma^{s}_{\mu\mu'}({\bf k, k'; q})=\int d{\bf r} \psi^{\dag}_{\mu' {\bf k'}}({\bf r})V_{s {\bf q}}({\bf r}) \psi_{\mu {\bf k}}({\bf r})
\end{eqnarray}
where the electron wave functions for the $K$ point are $\psi_{\mu {\bf k}}({\bf r})=\left(\mu~ e^{i \phi_{\bf k}} \right)^{T} \exp(i{\bf k}{\bf r})/\sqrt{\cal{A}}$ with ${\cal A}$ the normalizing area and $\phi_{\bf k}$ the polar angle of the vector ${\bf k}$.
%\begin{equation}\psi_{\mu {\bf k}}({\bf r})=\frac{1}{\sqrt{\cal{A}}} \left( \begin{array}{cc} \mu \\ e^{i \phi_{\bf k}}  \end{array} \right)\exp(i{\bf k}{\bf r})~.\end{equation}
The perturbation of the graphene lattice potential created by a single ${s{\bf q}}$ phonon mode is given by $V_{s{\bf q}}({\bf r})= \alpha v_{F} V_{s}({\bf q}) \exp(i{\bf q}{\bf r})/\sqrt{\cal{A}}$
%\begin{equation}V_{s{\bf q}}({\bf r})= \alpha \frac{\hbar v_{F} }{\sqrt{\cal{A}}} V_{s}({\bf q}) \exp(i{\bf q}{\bf r})\end{equation}
with the interaction matrices represented as \cite{ando}
\begin{equation}
V_{s}({\bf q})= \left( \begin{array}{cc} 0 & i^{1+s}e^{-i \phi_{\bf q}} \\ i^{1-s} e^{i \phi_{\bf q}} & 0 \end{array} \right)
\end{equation}
for the longitudinal ($s=1$) and transverse  ($s=0$) phonons. % at the ${\bf \Gamma}$ point of the graphene band.
The dimensionless coupling constant is defined as $\alpha =\beta/b^{2} \sqrt{2\bar{\sigma}\omega_{s \bf q}}$
%\begin{equation}
%\alpha =\frac{\beta}{b^{2}} \sqrt{\frac{\hbar}{2\bar{\sigma}\omega_{s}(q)}}
%\end{equation}
where $\bar{\sigma}$ is the surface mass density unit cells, $\beta\sim 2$ a dimensionless tight-binding parameter, $b$ the bond length between adjacent carbon atoms. This yields weak coupling with $\alpha^{2}\sim 0.02$ \cite{tse2}.

%For the $K'$ point\begin{equation}\psi_{\mu {\bf k}}({\bf r})=\frac{1}{\sqrt{\cal{A}}} \left( \begin{array}{cc} e^{i \phi_{\bf k}} \\  \mu  \end{array} \right)\exp(i{\bf k}{\bf r})\end{equation}and the interaction matrices are obtained from the relation $V^{K'}_{\alpha}({\bf q})=V^{K}_{\alpha}({-\bf q})^{*}$ taking into account that $V_{\alpha}({\bf q})^{\dag}=V_{\alpha}({-\bf q})$. 

Further we focus only on the part of the spectrum near the threshold of longitudinal optical phonon emission by electrons, $\e_{c}=+\omega_{0}$. The singular behavior of the spectrum near the threshold of phonon emission by Fermi holes, $\e_{c}=-\omega_{0}$, as well as for the transverse optical phonons can be treated independently in a similar way.

In the energy region of our interest, $\e\approx \e_{c}$, we can make several simplifications using the threshold approximation. In equations corresponding to the diagrams of Fig.~\ref{fg2} we take the electron Green functions as retarded and after the integration over the phonon energetic parameter $\omega$ replace it by $\omega_{0}$ in all internal electron lines. In proximity of the threshold in the conductance band, the leading contributions to the summation over the chirality of the internal electron lines make the {\it singular terms} with $\bar{\mu}=+1$ and $\mu'=+1$. The energetic parameter of the exact Green functions in the dangerous intersections in the second terms in the rhs of the equations in Fig.~\ref{fg2}, $\e-\omega_{0}$, lies far from the threshold $\e_{c}$ where perturbation theory is applicable. Therefore, the internal exact Green functions can be replaced by the bare function $G_{0+}(\e-\omega_{0},{\bf k})$.
\begin{figure*}[t]
\includegraphics[height=.305\linewidth]{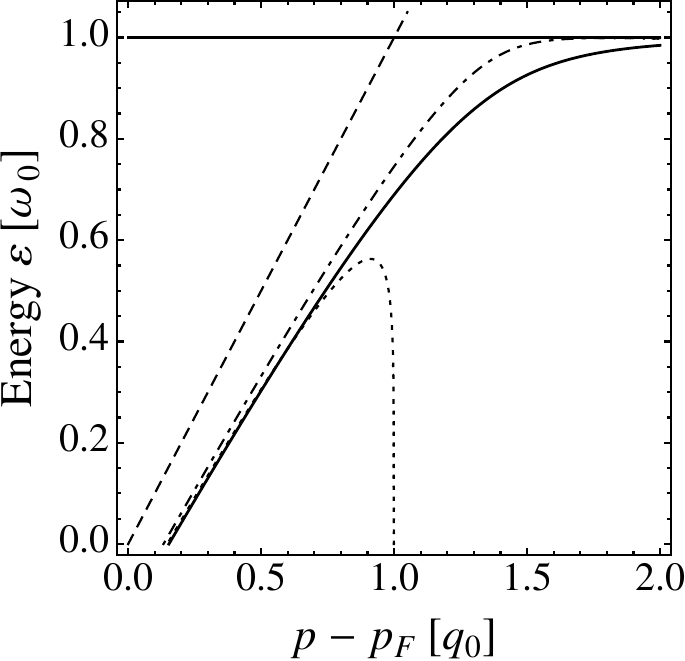}\includegraphics[height=.32\linewidth]{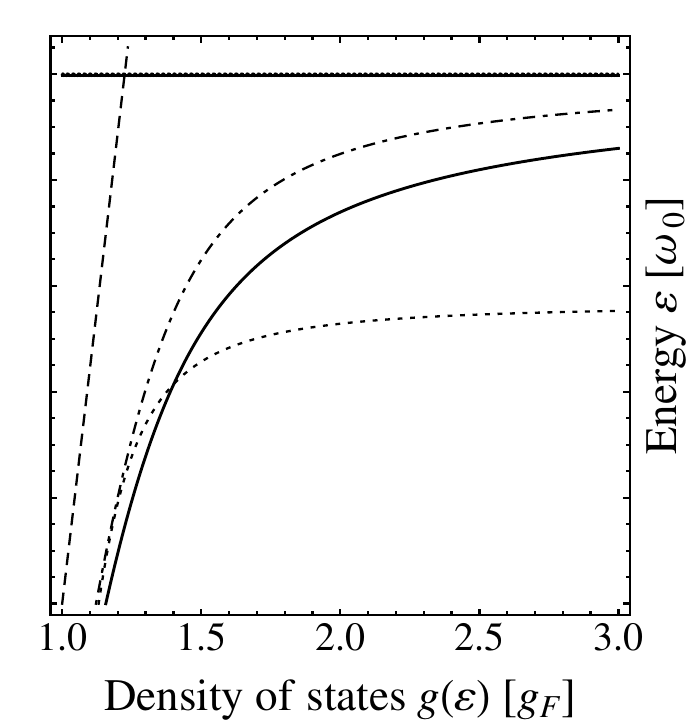}\hspace{10mm}\includegraphics[height=.301\linewidth]{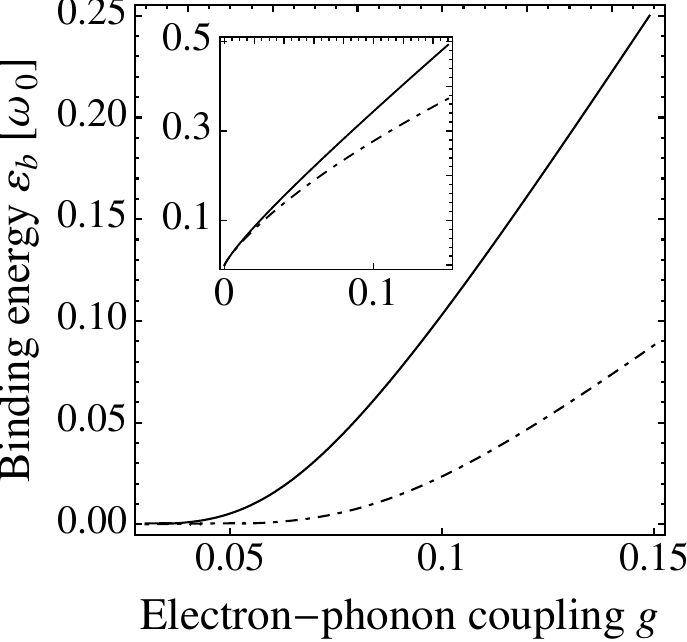}
\caption{(right) The energy spectrum of the electron-optical phonon quasiparticle for the electron density $n=5.6\times10^{13}$ cm$^{-2}$. The horizontal solid line is the phonon emission threshold. (mid) The corresponding density of states $g(\e)$ of the quasiparticle in units of $g_{F}=2k_{F}/\pi v_{F}$. (left) The electron-phonon binding energy versus the bare coupling $g$ for $p-p_{F}=1.5q_{0}$. Inset shows the binding energy of the hybrid states at the resonance $p-p_{F}=q_{0}$. In all figures the dashed lines represent the bare Dirac fermions, the dotted and dot-dashed curves are calculated, respectively, within the Rayleigh-Schr\"odinger and Wigner-Brillouin perturbative approaches. The solid curve is obtained within the present theory, taking into account the singular vertex corrections beyond perturbation theory. } 
\label{fg3}
\end{figure*}
Furthermore, in all integrations over the absolute values of the electron momenta corresponding to the dangerous intersections in Fig.~\ref{fg2}, only the small regions near the Fermi wave vector, $|{\bf k}|\approx k_{F}$ and $|{\bf k'}|\approx k_{F}$, give the main contribution to the integrals in these equations. Therefore, in this approximation one can take the quantities $\gamma$, $\square$, and $\Gamma$ out of the integrations over $k$ and $k'$. This allows us write the Dyson equation corresponding to Fig.~\ref{fg2} as
\begin{eqnarray}
G^{-1}_{\mu}(\e,{\bf p})&=&G^{-1}_{0 \mu}(\e,{\bf p}) - \Lambda_{+}(\e) \int_{0}^{2\pi}d \phi_{\bf k} \label{dyseq}\\
&\times&
\gamma_{\mu +}({\bf p, k_{F}; k_{F}-p})\Gamma_{+\mu}(\e | {\bf k_{F}, p; p-k_{F}}) \nonumber
\end{eqnarray}
and represent the equation for the nonanalytical vertex part in the following way
\begin{eqnarray}
&&\Gamma_{+\mu}(\e | {\bf k_{F}, p; p-k_{F}})=\gamma_{+\mu}({\bf k_{F}, p; p-k_{F}})  \label{threevertex} \\
&+&
\Lambda_{+}(\e) \int_{0}^{2\pi}d \phi_{\bf k'} \square_{++}(\e | {\bf k_{F}, p-k_{F}; k'_{F}, k'_{F}-p}) \nonumber \\
&\times& 
\Gamma_{+\mu}(\e | {\bf k'_{F}, p; p-k'_{F}})~. \nonumber
\end{eqnarray}
Here we introduce the following singular function
\begin{eqnarray}
\Lambda_{+}(\e)&=&\sum_{\nu,{\bf k}}G_{0\nu}(\e-\omega_{0},{\bf k}) \nonumber\\
&=&
\frac{{\cal A}}{2\pi}\int \frac{\left(1-\theta(\e_{+{\bf k}}-\ef)\right) k d k}{\e+\e_{F}-\omega_{0}-v_{F}{k}+i 0}
\label{lambda}
\end{eqnarray}
where the unit step $\theta(x)$ function is what remains from the Fermi functions at $T=0$. The divergence of the integral at large values of $k$ is related to the linearity of the graphene energy band. The cutoff of the integral at large momenta of the order of the inverse lattice constant contributes to the regular part of the integral. We are interested in its singular part due to the phonon emission threshold. It comes from the low limit of the integral, {\it i.e.} from the momentum range close to the Fermi wave vector $k_{F}$, which gives
\begin{eqnarray}
\Lambda_{+}(\e)&\propto& - \frac{{\cal A}}{2\pi} \frac{k_{F}}{v_{F}}\ln{\frac{{\e}_{F}}{\omega_{0}-\e}}~.
\end{eqnarray}
%Note that $\Lambda_{+}(\e)$ and the lowest order mass operator $\Sigma(\e)$ \cite{calandra,tse2,stauber} have the same singular behavior at $\e=\omega_{0}$.
Because the energetic parameter of the electron internal Green function $G_{0\mu''}(\e-2\omega_{0})$ in the four vertex part $\square_{++}(\e )$ is far from the threshold $\e_{c}$ (the four vertex part has no dangerous intersection along one electron and one phonon lines), $\square_{++}(\e )$ can be expanded with respect to $\alpha$ and be replaced by the simplest diagram shown in Fig.~\ref{fg1}(d). We restrict ourselves to highly doped samples where the Fermi energy is larger than the phonon energy. In this regime the important contribution to the four vertex part $\square_{++}(\e)$ comes from scattered virtual phonons with approximately equal antiparallel momenta ${\bf q= p-k}$ and ${\bf q'=p-k'}$ and we replace $\e_{\bf p-q'-q}$ by $\e_{F}$ in the electron Green function and retain only the term with chirality $\mu''=+1$ in the sum corresponding to the internal electron line. For such dominant scattering events the vertex parts $\gamma$ in $\square_{++}(\e)$ depend only on ${\bf q}$ or ${\bf q'}$ and 
the four vertex part becomes decoupled as
\begin{eqnarray}
\square_{++}(\omega_{0}|{\bf q; q'})\approx G_{0+}(-\omega_{0},k_{F}) 
\gamma_{++}({\bf q}) \gamma_{++}({\bf -q'})~. 
\label{fourvertex}
\end{eqnarray}
Taking $\mu=+1$ in all electron external lines and introducing a new amplitude 
\begin{eqnarray}
\hat{\Gamma}(\e, {\bf p})= \int_{0}^{2\pi}d \phi_{\bf q} 
\gamma_{++}({\bf -q}) \Gamma_{++}(\e | {\bf p-q, p ; q})
\label{thvertex}
\end{eqnarray}
as well as the form factor 
\begin{eqnarray}
\Upsilon(p)= \int_{0}^{2\pi}d \phi_{\bf q} |\gamma_{++}({\bf q})|^{2}  \approx 2\pi \alpha^{2}\frac{v_{F}^{2}}{\cal{A}}~,  
\label{cff}
\end{eqnarray}
and with the help of Eqs.~(\ref{threevertex}), (\ref{fourvertex})--(\ref{cff}) we get %for the vertex part
\begin{eqnarray}
\hat{\Gamma}(\e, {\bf p})&=&\frac{\Upsilon(p)}{1-\Lambda_{+}(\e) G_{0+}(-\omega_{0},k_{F})
\Upsilon(p) }~. 
\label{hvertex}
\end{eqnarray}
Combining Eqs.~(\ref{dyseq}), (\ref{thvertex}), and (\ref{hvertex}) and making use of the explicit expressions for the respective functions in these equations, we derive the following dispersion relation
\begin{eqnarray}
\e -v_{F}(p-p_{F}) = -\omega_{0} \frac{ g\ln\frac{{\e}_{F}}{\omega_{0}-\e}}{1- g\ln\frac{{\e}_{F}}{\omega_{0}-\e}}~. 
\label{fde}
\end{eqnarray}
Here we introduce a renormalized electron-phonon coupling $g=\alpha^{2} \ef/\omega_{0}$. Eq.~(\ref{fde}) includes the singular vertex corrections beyond perturbation theory and results in new properties of the electron-phonon quasiparticle in graphene. The perturbative Wigner-Brillouin (WB) approach is recovered if the dominator in the rhs of Eq.~(\ref{fde}) is set equal to $1$. Within the Rayleigh-Schr\"odinger (RS) theory the exact energy in the rhs of Eq.~(\ref{fde}) additionally should be replaced by $\e=v_{F}(p-p_{F})$.
The corresponding single-particle density of states is given by $g(\e ) = \left. 4p/(2\pi \partial {\e} /\partial p)\right|_{{\e} = {\e}(p)} $ where $\e (p)$ is the solution of Eq.~(\ref{fde}).
 
In Fig.~\ref{fg3} we plot the spectrum characteristics of the electron-phonon quasiparticle, calculated from Eq.~(\ref{fde}). We consider only the energy range $0<\e<\omega_{0}$. Above the threshold, $\e>\omega_{0}$, there exists a continuum of decaying states and no true elementary excitations. For $\e<0$, the threshold $\e_{c}=-\omega_{0}$, which refers to the Fermi holes, has a similar effect on the spectrum. As seen in Fig.~\ref{fg3}(left) at small values of $p-p_{F} < q_{0}$  ($q_{0}=\omega_{0}/v_{F}$) the effect of vertex corrections is weak and the spectrum obtained from perturbative RS and WB theories provides an adequate description. In this limit the average number of phonon states $N\sim g$ and the density of states is linear with $\e$. At $\e=0$ the shift of $g(\e)$ from $g_{F}$, the density of states of the bare Dirac fermions at the Fermi level, is proportional to the coupling $g$ (Fig.~\ref{fg3}(mid)). With an increase of the momentum $p$ the RS approach fails completely while the spectrum, obtained within the two other approaches for $p-p_{F}\sim q_{0}$, describes the electron-phonon hybrid states  with $N\sim 1/2$. It is seen in the inset in Fig.~\ref{fg3}(left) that the vertex corrections increase the binding energy, $\e_{b}=\omega_{0}-\e$, but the effect is still relatively modest. 

For even larger momenta $p-p_{F}> q_{0}$ the spectrum obtained within WB perturbation theory (this approach actually has been used in Refs.~\onlinecite{park,tse2,calandra,stauber}) converge asymptotically to the phonon energy when $p\rightarrow \infty$ hence does not support an electron-phonon bound state. In contrast, the vertex corrections become especially important in this region and open a small gap under the threshold. The true spectrum obtained from Eq.~(\ref{fde}) always remains below the phonon energy and corresponds to the electron-phonon {\it bound state} with $N\sim 1$. The velocity of the bound state tends to zero while the density of states increases strongly with $p$. 
Although the binding energy of the bound state remains finite for $p\rightarrow \infty$, it exhibits a stark exponential dependence on the coupling constant, $\e^{\infty}_{b}={\e}_{F}\exp(-1/g)$, and with its sub-Kelvin value for experimentally accessible values of $g$ is hardly measurable. At finite momenta $p-p_{F}> q_{0}$, we find, however, that the singular vertex corrections increase strongly the binding energy $\e_{b}$ in comparison with that obtained within the perturbative WB approach (see Fig.~\ref{fg3}(left)). For $p-p_{F}=1.5 q_{0}$ and for the doping level $n=5.6\times10^{13}$ cm$^{-2}$ (corresponding to $g\approx 0.09$) 
we find for the binding energy $\e_{b}\approx14.5$ meV. It is about a factor of $5$ larger than the corresponding WB perturbative value and this difference increases strongly with $g$ (Fig.~\ref{fg3}(left)). 
This enhancement results in a significant deviation from the linear Dirac spectrum that should manifest itself in angle-resolved measurements with the resolution smaller than $10$ meV as stark delta-function peaks at frequencies $\omega_{0}-\e_{b}$ and wave vectors larger than $q_{0}$.

In conclusion, we have calculated the fine structure of the Dirac spectrum in graphene in the proximity of the phonon emission threshold. The renormalized spectrum in the immediate neighborhood below the threshold corresponds to the electron-phonon bound state. Our calculations result in a strong enhancement of the electron-phonon binding energy due to the singular vertex corrections, which can be probed in experiment.

We thank E. Rashba for his useful comments and acknowledge support from the Belgium Science Policy (IAP) and BELSPO.

\end{document}